\documentstyle[prd,aps,eqsecnum]{revtex}
\input epsf
\begin{document}
\preprint{PU-RCG/98-21, astro-ph/9812204}
\draft
\renewcommand{\topfraction}{0.8}
\renewcommand{\bottomfraction}{0.8}
%
\title{Dynamics of Assisted Inflation} 
\author{Karim A. Malik and David Wands}
\address{School of Computer Science and Mathematics, University of 
Portsmouth, Portsmouth PO1 2EG,~~~U.~K.}  
\date{\today} 
\maketitle
\begin{abstract}
We investigate the dynamics of the recently proposed model of assisted
inflation. In this model an arbitrary number of scalar fields with 
exponential potentials evolve towards an inflationary scaling solution,
even if each of the individual potentials is too steep to support
inflation on its own.
By choosing an appropriate rotation in field space we can write down
explicitly the  potential for the weighted mean field along the
scaling solution and for fields orthogonal to it.  
This demonstrates that the potential has a global minimum along 
the scaling solution. 
We show that the potential close to this attractor in the
rotated field space is analogous to a hybrid inflation model, but with
the vacuum energy having an exponential dependence upon a dilaton
field. 
We present analytic solutions describing homogeneous and
inhomogeneous  perturbations about the attractor solution without
resorting to slow-roll approximations. We discuss the curvature and 
isocurvature perturbation spectra produced from vacuum fluctuations 
during assisted inflation.
\end{abstract}
\pacs{PACS number: 98.80.Cq \hfill Preprint  PU-RCG/98-21,
astro-ph/9812204} 
%
 

\section{Introduction}

A single scalar field with an exponential potential is known to drive
power-law inflation, where the cosmological scale factor grows as
$a\propto t^p$ with $p>1$, for sufficiently flat
potentials~\cite{PL,Halliwell,BB,CLW}.  Liddle, Mazumdar and Schunck
\cite{LMS} recently proposed a novel model of inflation driven by
several scalar fields with exponential potentials. Although each
separate potential,
\begin{equation}
\label{Vi}
V_i = V_0 \exp \left( - \sqrt{16\pi\over p_i} {\phi_i\over m_{\rm Pl}}
\right) \,,
\end{equation}
may be too steep to drive inflation by itself ($p_i<1$), the combined
effect of several such fields, with total potential energy
\begin{equation}
V = \sum_{i=1}^n V_i \,,
\end{equation}
leads to a power-law expansion $a\propto
t^{\bar{p}}$ with~\cite{LMS}
\begin{equation}
\label{barp}
\bar{p} = \sum_{i=1}^n p_i\,,
\end{equation}
provided $\bar{p}>1/3$.
Supergravity theories typically predict steep exponential
potentials, but if many fields can cooperate to drive inflation, this
may open up the possibility of obtaining inflationary solutions in
such models.

Scalar fields with exponential potentials are known to possess
self-similar solutions in Friedmann-Robertson-Walker
models either in vacuum~\cite{Halliwell,BB} or in the presence of
a barotropic fluid~\cite{Wetterich,WCL,CLW,vdH}. In the presence of other
matter, the scalar field is subject to additional friction, due to the
larger expansion rate relative to the vacuum case. This means that a
scalar field, even if it has a steep (non-inflationary) potential may
still have an observable dynamical effect in a radiation or matter
dominated era~\cite{Wetterich95,Joyce,LV}.

The recent paper of Liddle, Mazumdar and Schunck~\cite{LMS} was the
first to consider the effect of additional scalar fields with
independent exponential potentials. They considered $n$ scalar fields
in a spatially flat Friedmann-Robertson-Walker universe with scale
factor $a(t)$. The Lagrange density for the fields is
\begin{equation}
{\cal L} = \sum_{i=1}^n - {1\over2} \left(\nabla\phi_i\right)^2 - V_i \,,
\end{equation}
with each exponential potential $V_i$ of the form given in Eq.~(\ref{Vi}). 
The cosmological expansion rate is then given by
\begin{equation}
\label{constraint}
H^2 = {8\pi \over 3m_{\rm Pl}^2} \sum_{i=1}^n \left( V_i + {1\over2}
\dot\phi_i^2 \right) \,,
\end{equation}
and the individual fields obey the field equations
\begin{equation}
\label{phieom}
\ddot\phi_i +3H\dot\phi_i = - {dV_i \over d\phi_i} \,.
\end{equation}

One can then obtain a scaling solution of the form~\cite{LMS}
\begin{equation}
{\dot\phi_i^2 \over \dot\phi_j^2} = {V_i \over V_j} = C_{ij} \,.
\end{equation}
Differentiating this expression with respect to time, and using the
form of the potential given in Eq.~(\ref{Vi}) then implies that
\begin{equation}
{1\over \sqrt{p_i}} \dot\phi_i - {1\over \sqrt{p_j}} \dot\phi_j = 0 \,,
\end{equation}
and hence
\begin{equation}
C_{ij} = {p_i \over p_j} \,.
\end{equation}
The scaling solution is thus given by~\cite{LMS}
\begin{equation}
\label{attractor}
{1\over\sqrt{p_i}} \phi_i - {1\over\sqrt{p_j}} \phi_j 
 = {m_{\rm Pl} \over \sqrt{16\pi}} \ln {p_j\over p_i} \,.
\end{equation}
A numerical solution with four fields is shown in Fig.~1 as an example.
In Ref.\cite{LMS} the authors demonstrated the existence of a scaling
solution for $n$ scalar fields written in terms of a single re-scaled
field $\tilde\phi=\sqrt{\bar{p}/p_1}\phi_1$. The choice of $\phi_1$
rather than any of the other fields is arbitrary as along the
scaling solution all the $\phi_i$ fields are proportional to one
another.

\begin{figure}[t]
\centering 
\leavevmode\epsfysize=5cm \epsfbox{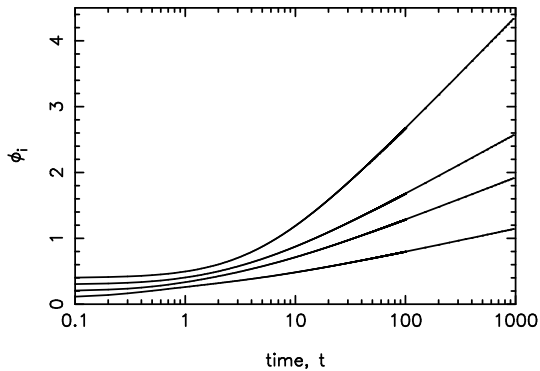}\\ 
\caption[Phi fields]{\label{Fone}
Evolution of four fields $\phi_1$, $\phi_2$, $\phi_3$ and $\phi_4$
(from bottom to top) during assisted inflation 
with $p_1=0.3$, $p_2=1$, $p_3=2$ and $p_4=7$.} 
\end{figure}

In this paper we will prove that this scaling solution is the
late-time attractor by choosing a redefinition of fields (a rotation
in field space) which allows us to write down the effective potential
for field variations orthogonal to the scaling solution and show that
this potential has a global minimum along the attractor solution.  In
general the full expression for an arbitrary number of fields is
rather messy so we first give, in Sect.~\ref{sect2}, the simplest case
where there are just two fields, and then extend this to $n$ fields in
Sect.~\ref{sectn}.  The resulting inflationary potential is similar to
that used in models of hybrid inflation and we show in
Sect.~\ref{secthy} that assisted inflation can be interpreted as a
form of ``hybrid power-law inflation''. As in the case of power-law or
hybrid inflation, one can obtain analytic expressions for
inhomogeneous linear perturbations close to the attractor trajectory
without resorting to slow-roll type approximations. Thus we are able
to give exact results for the large-scale perturbation spectra due to
vacuum fluctuations in the fields in Sect.~\ref{sectpert}. We discuss
our results in Sect.~\ref{sectdisc}.


\section{Two field model}
\label{sect2}

We will restrict our analysis initially to just two scalar fields, 
$\phi_1$ and $\phi_2$, with the Lagrange density
\begin{equation}
\label{V2}
{\cal L} = -{1\over2}(\nabla\phi_1^2) - {1\over2}(\nabla\phi_2^2) - 
V_0 \left[ \exp \left( - \sqrt{16\pi\over p_1} {\phi_1\over m_{\rm
Pl}} \right)
 + \exp \left( - \sqrt{16\pi\over p_2} {\phi_2\over m_{\rm Pl}}
\right) \right]
\,.
\end{equation}

We define the fields
\begin{eqnarray}
\label{barphi2}
\bar\phi_2 &=& { \sqrt{p_1} \phi_1 + \sqrt{p_2} \phi_2 \over \sqrt{p_1 +
p_2} } + {m_{\rm Pl} \over \sqrt{16\pi(p_1+p_2)}} \left( p_1 \ln
{p_1\over p_1+p_2} + p_2 \ln {p_2 \over p_1+p_2} \right)  \,,\\
\label{barsigma2}
\bar\sigma_2 &=& { \sqrt{p_2} \phi_1 - \sqrt{p_1} \phi_2 \over \sqrt{p_1 +
p_2} } + {m_{\rm Pl} \over \sqrt{16\pi}} \sqrt{p_1p_2 \over p_1+p_2}
\ln {p_1 \over p_2} \,,
\end{eqnarray}
to describe the evolution along and orthogonal to the scaling
solution, respectively, by applying a Gram-Schmidt orthogonalisation
procedure.

The re-defined fields $\bar\phi_2$ and $\bar\sigma_2$ are 
orthonormal linear combinations of the original fields $\phi_1$ and $\phi_2$.
They represent a rotation, and arbitrary shift of the origin, in
field-space. Thus $\bar\phi_2$ and $\bar\sigma_2$ have canonical kinetic 
terms, and the Lagrange density given in Eq.~(\ref{V2}) can be written as
\begin{equation}
\label{L2}
{\cal L} = -{1\over2}(\nabla\bar\phi_2^2) - {1\over2}(\nabla\bar\sigma_2^2) 
- \bar{V}(\bar\sigma_2) \exp \left( - \sqrt{16\pi\over p_1+p_2}
{\bar\phi_2\over m_{\rm Pl}} \right) 
\,,
\end{equation}
where
\begin{equation}
\label{barV2}
\bar{V}(\bar\sigma_2) = V_0 \left[
 {p_1\over p_1+p_2}
 \exp \left( -\sqrt{16\pi \over p_1+p_2} \sqrt{p_2\over p_1}
  {\bar\sigma_2\over m_{\rm Pl}} \right) + {p_2\over p_1+p_2} 
 \exp \left( \sqrt{16\pi \over p_1+p_2} \sqrt {p_1\over p_2}
 {\bar\sigma_2\over m_{\rm Pl}} \right) \right] \,.
\end{equation}
It is easy to confirm that $\bar{V}(\bar\sigma_2)$ has a global minimum 
value $V_0$ at $\bar\sigma_2=0$, which  implies that $\bar\sigma_2=0$ is 
the late time attractor, which coincides with the scaling solution given in
Eq.~(\ref{attractor}) for two fields.

Close to the scaling solution we can expand about the minimum, to
second-order in $\bar\sigma_2$, and we obtain
\begin{equation}
\label{V2Taylor}
V(\bar\phi_2,\bar\sigma_2) \approx
 V_0 \left[ 1 + {8\pi \over (p_1+p_2)} {\bar\sigma_2^{~2}\over m_{\rm Pl}^2} 
\right] \exp \left( - \sqrt{16\pi\over p_1+p_2} {\bar\phi_2\over m_{\rm Pl}}
 \right) \,. 
\end{equation}

Note that the potential for the field $\bar\sigma_2$ has the same form as in
models of hybrid inflation~\cite{hybrid1,hybrid2} where the inflaton field
rolls towards the minimum of a potential with non-vanishing potential
energy density ${V}_0$. Here there is in addition a ``dilaton'' field,
$\bar\phi_2$, which leads to a time-dependent potential energy density
as $\bar\sigma_2\to0$. 
Assisted inflation is related to hybrid
inflation~\cite{hybrid1,hybrid2} in the
same way that extended inflation~\cite{extinf} was related to Guth's
old inflation model~\cite{Guth}.
As in hybrid or extended inflation, we require a phase transition to bring
inflation to an end. Otherwise the potential given by
Eq.~(\ref{V2Taylor}) leads to inflation into the indefinite future.


\section{many field model}
\label{sectn}

We will now prove that the attractor solution presented in
Ref.~\cite{LMS} is the global attractor for an arbitrary number of
fields with exponential potentials of the form given in
Eq.~(\ref{Vi}), using proof by induction. To do this, we recursively
construct the orthonormal fields and their potential.

Let us assume that we already have $n$ fields $\phi_i$ with
exponential potentials $V_i$ of the form given in Eq.~(\ref{Vi}) and
that it is possible to pick $n$ orthonormal fields
$\bar\sigma_2,\ldots,\bar\sigma_n$ and $\bar\phi_n$ such that the sum
of the individual potentials $V_i$ can be written as
\begin{equation}
\label{Vsumn}
\sum_{i=1}^n V_i
= \bar{V}_{n} \exp \left( - \sqrt{16\pi\over \bar p_{n}}
{\bar\phi_{n}\over m_{\rm Pl}} \right)  \,,
\end{equation}
where we will further assume that $\bar{V}_n= \bar{V}_n
(\bar\sigma_i)$ has a global minimum $\bar V_n(0)=V_0$ when
$\bar\sigma_i=0$ for all $i$ from $2$ to $n$.

It is possible to extend this form of the potential
to $n+1$ fields if we consider an additional field $\phi_{n+1}$ with
an exponential potential $V_{n+1}$ of the form given in
Eq.~(\ref{Vi}). 
Analogously to the two field case, we define
\begin{eqnarray}
\label{barphin+1}
\bar\phi_{n+1} &=& { \sqrt{\bar p_n} \bar\phi_n + \sqrt{p_{n+1}}
\phi_{n+1} \over \sqrt{\bar p_{n+1} } } + {m_{\rm Pl} \over
\sqrt{16\pi(\bar p_{n+1})}} \left( \bar 
p_{n} \ln {\bar p_n\over \bar p_{n+1}} + p_{n+1} \ln {p_{n+1} \over
\bar p_{n+1}} \right) \,,\\
\bar\sigma_{n+1} &=& { \sqrt{p_{n+1}} \bar\phi_n - \sqrt{\bar p_n}
\phi_{n+1} \over \sqrt{\bar p_{n+1}} } + {m_{\rm Pl} \over
\sqrt{16\pi}} \sqrt{\bar p_n p_{n+1} \over 
\bar p_{n+1} } \ln {\bar p_n \over p_{n+1} } \,,
\end{eqnarray}
where
\begin{equation}
\label{recurbarp}
\bar p_{n+1} = \bar p_n + p_{n+1} \ .
\end{equation}
Using these definitions we can show that the sum of the $n+1$
individual potentials $V_i$ can be written as
\begin{equation}
\sum_{i=1}^{n+1} V_i
 = \bar{V}_{n+1} \exp \left( - \sqrt{16\pi\over \bar p_{n+1}}
{\bar\phi_{n+1}\over m_{\rm Pl}} \right)  \,,
\end{equation}
where $\bar{V}_{n+1}=\bar{V}_{n+1}(\bar\sigma_i)$ is given by
\begin{equation}
\label{Vbarn+1}
\bar V_{n+1} = \bar V_n 
 {\bar p_n \over \bar p_{n+1} }
 \exp \left( -\sqrt{16\pi \over \bar p_{n+1}} \sqrt{p_{n+1}\over \bar p_n}
  {\bar\sigma_{n+1} \over m_{\rm Pl}} \right) + V_0~ {p_{n+1}\over
 \bar p_{n+1}}  
 \exp \left( \sqrt{16\pi \over \bar p_{n+1}} \sqrt {\bar p_n\over p_{n+1}}
 {\bar\sigma_{n+1} \over m_{\rm Pl}} \right) \,.
\end{equation}
Because we have assumed that $\bar{V}_n$ has a global minimum value 
$\bar{V}_n(0)=V_0$ when $\bar\sigma_i=0$ for all $i$ from $2$ to $n$, 
one can verify that $\bar{V}_{n+1}$ also has a
minimum value $\bar{V}_{n+1}(0)=V_0$ when $\bar\sigma_i=0$, for all
$i$ from $2$ to $n+1$.

However, we have already shown in Sect.~\ref{sect2} that for two 
fields $\phi_1$ and $\phi_2$, we can define two fields $\bar\phi_2$ 
and $\bar\sigma_2$,
given in Eqs.~(\ref{barphi2}) and~(\ref{barsigma2}) whose combined
potential given in Eq.~(\ref{L2}) is of the form required in
Eq.~(\ref{Vsumn}), with $\bar{p}_2=p_1+p_2$. Hence we can write the
potential in the form given in Eq.~(\ref{Vsumn}) for $n$ fields, for all
$n\geq2$, with 
\begin{equation}
\label{newbarp}
\bar p \equiv \bar p_n = \sum_{i=1}^n p_i \ .
\end{equation}
Equations (\ref{barphi2}) and (\ref{barphin+1}) then lead us to the
non-recursive expression for the ``weighted mean field''
\begin{equation}
\bar\phi \equiv \bar\phi_n = \sum_{i=1}^n \left( \sqrt{ { p_i \over
\bar{p} } }  
\phi_i + {m_{\rm Pl} \over \sqrt{16\pi\bar{p}}} ~p_i \ln {p_i\over \bar{p}}
 \right) \,,
\end{equation}
which describes the evolution along the scaling solution.
This is simply a rotation in field space plus an arbitrary shift, chosen
to preserve the form of the potential given in Eq.~(\ref{Vsumn}).
The $n-1$ fields $\bar\sigma_i$ describe the evolution orthogonal to
the attractor trajectory.

The potential $\bar{V}_n$ has a global minimum at $\bar\sigma_i=0$, which
demonstrates that this is the stable late-time attractor.
{}From Eqs.~(\ref{barV2}) and~(\ref{Vbarn+1}) we get a closed
expression for $\bar{V}_n$, 
\begin{eqnarray}
\label{Vbarn}
\bar{V}_n &=& V_0 \left\{ \frac{p_1}{\bar p} \exp \left[
-\frac{\sqrt{16\pi}}{m_{\rm Pl}} ~\sum^{n}_{i=2}
\sqrt{\frac{p_{i}}{\bar p_{i} \bar p_{i-1}} } ~\bar \sigma_i \right] 
+~\sum^{n-1}_{i=2} \frac{p_{i}}{\bar p} \exp \left[
\frac{\sqrt{16\pi}}{m_{\rm Pl}} \left( 
\sqrt{\frac{\bar p_{i-1}}{\bar p_{i} p_{i}} } ~\bar \sigma_i
- \sum^{n}_{j=i+1} \sqrt{\frac{ p_{j}}{\bar p_{j} \bar p_{j-1}} }
~\bar \sigma_j \right) \right] \right. \nonumber \\ 
&& \qquad \left. + ~ \frac{p_n}{\bar p} \exp \left[
\frac{\sqrt{16\pi}}{m_{\rm Pl}} \sqrt{\frac{\bar p_{n-1}}{\bar p
p_{n}} } ~\bar \sigma_n  
\right]  \right\} .
\end{eqnarray}
Close to the attractor trajectory (to second order in $\bar\sigma_i$)
we can write a Taylor expansion for the potential
\begin{equation}
\label{VnTaylor}
\sum_{i=1}^n V_i
= V_0 \left( 1 +  {8\pi \over {\bar{p} m_{\rm Pl}^2}} \sum_{j=2}^{n}
 \bar\sigma_j^2  
\right) \exp \left( - \sqrt{16\pi\over \bar p}
{\bar\phi_{n}\over m_{\rm Pl}} \right)  \,.
\end{equation}
Note that this expression is dependent only upon $\bar p$ and not on the 
individual $p_i$.


\section{Stringy hybrid inflation}
\label{secthy}

The form of the potentials in Eqs.~(\ref{V2Taylor})
and~(\ref{VnTaylor}) is reminiscent of the effective potential
obtained in the Einstein conformal frame from Brans-Dicke type
gravity theories.  
The appearance of the weighted mean field, $\bar\phi$, as a
``dilaton'' field in the potential suggests that the matter Lagrangian
might have a simpler form in a conformally related frame. 
If we work in terms of a conformally re-scaled metric
\begin{equation}
\label{tildeg}
\tilde{g}_{\mu\nu} = \exp \left( - \sqrt{ 16\pi \over \bar{p}
} {\bar\phi \over m_{\rm Pl}} \right) g_{\mu\nu} \,,
\end{equation}
then the Lagrange density given in Eq.~(\ref{L2}) becomes
\begin{equation}
\label{confL}
\tilde{{\cal L}} = \exp \left( \sqrt{ 16\pi \over \bar{p} }
 {\bar\phi \over m_{\rm Pl}} \right) 
  \times \left\{ - {1\over2}
 \left(\tilde\nabla\bar\phi\right)^2 - \sum_{i=2}^{n} {1\over2}
 \left(\tilde\nabla\bar\sigma_i\right) - \bar{V} \right\} \,,
\end{equation}

In this conformal related frame the field $\bar\phi$ is non-minimally
coupled to the gravitational part of the Lagrangian.  The original
field equations were derived from the full action, including the
Einstein-Hilbert Lagrangian of general relativity,
\begin{equation}
S = \int d^4x \sqrt{-g} \left[ {m_{\rm Pl}^2 \over 16\pi} R + {\cal L}
\right] \,,
\end{equation}
where $R$ is the Ricci scalar curvature of the metric $g_{\mu\nu}$. In
terms of the conformally related metric given in Eq.~(\ref{tildeg})
this action becomes (up to boundary terms~\cite{Wands94})
\begin{equation}
S = \int d^4x \sqrt{-\tilde{g}} e^{-\Phi} \left[ 
{m_{\rm Pl}^2 \over 16\pi} \tilde{R}  - \omega (\tilde\nabla\Phi)
-{1\over2} \sum_{i=1}^{n-1} (\tilde\nabla\bar\sigma_i)^2 - \bar{V}
\right] \,,
\end{equation}
where we have introduced the dimensionless dilaton field
\begin{equation}
\Phi = - \sqrt{16\pi \over \bar{p}} {\bar\phi \over m_{\rm Pl}} \,,
\end{equation}
and the dimensionless Brans-Dicke parameter
\begin{equation}
\omega = {\bar{p} - 3 \over 2} \,.
\end{equation}

Thus the assisted inflation model is identical to $n-1$ scalar fields
$\bar\sigma_i$ with a hybrid inflation type potential
$\bar{V}(\bar\sigma_i)$ in a string-type gravity theory with dilaton,
$\Phi\propto\bar\phi$. However, we note that in order to obtain
power-law inflation with $\bar{p}\gg1$ the dimensionless constant
$\omega$ must be much larger than that found in the low-energy limit
of string theory where $\omega=-1$.


\section{Perturbations about the attractor}
\label{sectpert}

The redefined orthonormal fields and the potential allow us to give
the equations of motion for the independent degrees of freedom. If we
consider only linear perturbations about the attractor then the
energy density is independent of all the fields except $\bar\phi$, and
we can solve the equation for the $\bar \phi$ field 
analytically.

The field equation for the weighted mean field is 
\begin{equation}
\ddot{\bar\phi} + 3H\dot{\bar\phi} = \sqrt{{16\pi \over \bar p}} {V
\over m_{\rm Pl}} \,.
\end{equation}
Along the line $\bar\sigma_i=0$ for all $i$ in field space,
we have
\begin{equation}
V = V_0 \exp \left( - \sqrt{16\pi\over \bar{p}}
 {\bar\phi\over m_{\rm Pl}} \right) \,,
\end{equation}
and the well-known power-law solution~\cite{PL} 
with $a\propto t^{\bar{p}}$ is the late-time attractor~\cite{Halliwell,BB} 
for this potential, where
\begin{equation}
\bar\phi(t)=\bar\phi_{0} \ln\left(\frac{t}{t_0}\right),
\end{equation}
and $\bar\phi_{0}=m_{\rm{Pl}}\sqrt{\bar p/4\pi}$ and
$t_0=m_{\rm{Pl}}\sqrt{\bar p/8\pi V_0(3\bar p-1)}$.

\subsection{Homogeneous linear perturbations}

The field equations for the $\bar \sigma_i$ fields are
\begin{equation}
\ddot{\bar\sigma_i} + 3H\dot{\bar\sigma_i} + \frac{\partial
V}{\partial \bar\sigma_i}=0. 
\end{equation}
where the potential $V$ is given by Eqs.~(\ref{Vsumn}) and~(\ref{Vbarn}),
and the attractor solution corresponds to $\bar\sigma_i=0$.
Equation~(\ref{VnTaylor}) shows that we can neglect the back-reaction
of $\bar\sigma_i$ upon the energy density, and hence the cosmological
expansion, to first-order and the field equations have the solutions 
\begin{equation}
\label{sigmat}
\bar\sigma_i(t)=\Sigma_{i+}t^{s_{+}} + \Sigma_{i-}t^{s_{-}},
\end{equation}
where
\begin{equation}
s_{\pm}=\frac{3\bar p -1}{2}
 \left[-1 \pm \sqrt {\frac{3(\bar{p}-3)}{3\bar p -1}} ~ \right], 
\end{equation}
for $\bar{p}>3$, confirming that $\sigma_i=0$ is indeed a local
attractor. In the limit $\bar p \to\infty$ we 
obtain $s_i=-2$. For $1<\bar{p}<3$ the perturbations are under-damped 
and execute decaying oscillations about $\bar\sigma_i=0$.

\begin{figure}[t]
\centering 
\leavevmode\epsfysize=5cm \epsfbox{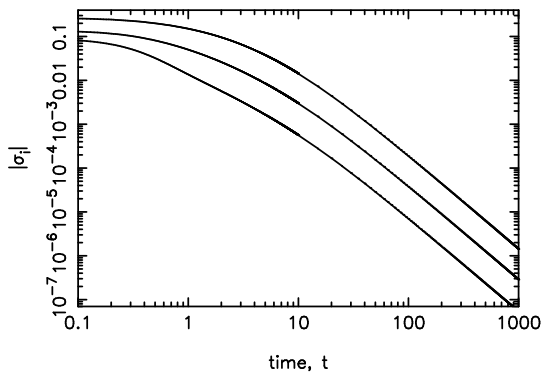}\\ 
\caption[Sigma fields]{\label{Ftwo}
Evolution of the fields $\bar\sigma_1$, $\bar\sigma_2$, and
$\bar\sigma_3$, orthogonal to the scaling solution, in the assisted
inflation model shown in Fig.~\ref{Fone}.}
\end{figure}

The form of the solutions given in Eq.~(\ref{sigmat}) for
$\bar\sigma_i(t)$ close to the attractor is the same for all the
orthonormal fields $\bar\sigma_i$, as demonstrated in Fig.~\ref{Ftwo}.
Their evolution is independent of the individual $p_i$ and determined
only by the sum, $\bar{p}$, as expected from the form of the
potential given in Eq.~(\ref{VnTaylor}).

\subsection{Inhomogeneous Linear Perturbations}

Conventional hybrid inflation and power-law inflation are two of the
very few models~\cite{LL93} in which one can obtain exact analytic
expressions for the spectra of vacuum fluctuations on all scales
without resorting to a slow-roll type approximation. In the case of
hybrid inflation, this is only possible in the limit that the inflaton
field $\sigma$ approaches the minimum of its potential and we can
neglect its back-reaction on the metric~\cite{GBW}. As the present
model is so closely related to both power-law and hybrid inflation
models close to the attractor, it is maybe not surprising then that we
can obtain exact expressions for the evolution of inhomogeneous linear
perturbations close to the scaling solution.

We will work in terms of the redefined fields $\bar\phi$ and
$\bar\sigma_i$, and their perturbations on spatially flat
hypersurfaces~\cite{MW}.  In the limit that $\bar\sigma_i\to0$
we can neglect the back-reaction of the $\bar\sigma_i$ field upon the
metric and the field $\bar\phi$.  Perturbations in the field
$\bar\phi$ then obey the usual equation for a single field driving
inflation~\cite{Mukhanov}, and perturbations in the field
$\bar\sigma_i$ evolve in a fixed background.  Defining
\begin{eqnarray}
u & = & a \delta\bar\phi \,,\\
v_i & = & a \delta\bar\sigma_i \,,
\end{eqnarray}
we obtain the decoupled equations of motion for perturbations with
comoving wavenumber $k$,
\begin{eqnarray}
u_k'' + \left( k^2 - {z'' \over z} \right) u_k &=& 0 \,,\\
v_{ik}'' + \left( k^2 + a^2 {d^2 V\over d\bar\sigma_i^2} - {a'' \over a}
\right) v_{ik} &=& 0 \,,
\end{eqnarray}
where~\cite{Mukhanov} $z\equiv a^2\bar\phi'/a'$ and a prime denotes
differentiation with respect to conformal time $\eta\equiv \int dt/a$.
For power-law expansion we have $z\propto a\propto
(-\eta)^{-\bar{p}/(\bar{p}-1)}$ and thus
\begin{equation}
{a'' \over a} = {\bar{p}(2\bar{p}-1) \over (\bar{p}-1)^2} ~ \eta^{-2}.
\end{equation}
We also have $aH\propto -\bar{p}/((\bar{p}-1)\eta)$ which gives
\begin{equation}
a^2 {d^2 V\over d\sigma^2}
 =\frac{ 2 (3 \bar{p}-1)}{(\bar p -1)^2} ~ \eta^{-2},
\end{equation}
where we have used the fact that $d^2V/d\bar\sigma_i^2=16\pi V/m_{\rm
Pl}^2$ along the attractor. 
The equations of motion therefore become
\begin{eqnarray}
u_k'' + \left( k^2 - {\nu^2 - (1/4) \over \eta^2} \right) u_k &=& 0 \,,\\
v_{ik}'' + \left( k^2 - {\lambda^2 - (1/4) \over \eta^2} \right)
v_{ik} &=& 0 \,,
\end{eqnarray}
where
\begin{eqnarray}
\nu &=& {3\over2} + {1\over \bar{p}-1} \,,\\
\lambda &=& {3\over2} {\sqrt{(\bar{p}-3)(\bar{p}-1/3)} \over \bar{p}-1}
\,,
\end{eqnarray}
and the general solutions in terms of Hankel functions are
\begin{eqnarray}
u_k = U_1 (-k\eta)^{1/2} H^{(1)}_\nu (-k\eta)
 + U_2 (-k\eta)^{1/2} H^{(2)}_\nu (-k\eta) \,, \\
v_{ik} = V_{1i} (-k\eta)^{1/2} H^{(1)}_\lambda (-k\eta)
 + V_{2i} (-k\eta)^{1/2} H^{(2)}_\lambda (-k\eta) \,.
\end{eqnarray}
Taking only positive frequency modes in the initial vacuum state for
$|k\eta|\gg1$ and normalising requires $u_k$ and $v_{ik}\to
e^{-ik\eta}/\sqrt{2k}$, which gives the vacuum solutions
\begin{eqnarray}
u_k &=& \frac{1}{2}(-\pi\eta)^{1/2} e^{\frac{\pi}{2}(\nu +1)i} 
H^{(1)}_\nu (-k\eta),  \\
v_{ik} &=& \frac{1}{2}(-\pi\eta)^{1/2} e^{\frac{\pi}{2}(\lambda +1)i} 
H^{(1)}_\lambda (-k\eta), 
\end{eqnarray}
In the opposite limit, i.e., $-k\eta \to 0$, we use the limiting form
of the Hankel functions, $H_{\nu}^{(1)}(z) \sim -(i/\pi)\Gamma (\nu)
z^{-\nu}$, and therefore on large scales, and at late times, we obtain
\begin{eqnarray}
u_k&\to& \frac{2^{\nu-1}}{\sqrt{\pi k}}e^{i\frac{\pi}{2}\nu} 
\left(-k\eta\right)^{\frac{1}{2}-\nu}\Gamma(\nu), \\
v_{ik}&\to& \frac{2^{\lambda-1}}{\sqrt{\pi k}}e^{i\frac{\pi}{2}\lambda} 
\left(-k\eta\right)^{\frac{1}{2}-\lambda}\Gamma(\lambda) \,.
\end{eqnarray}

The power spectrum of a Gaussian random field $\psi$ is conventionally
given by ${\cal P}_{\psi} \equiv \left(k^3/2\pi^2\right) \langle
|\psi|^2 \rangle$.  
The power spectra on large scales for the field perturbations
$\delta\bar\phi$ and $\delta\bar\sigma_i$ are thus
\begin{eqnarray}
{\cal P}_{\delta\phi}^{~~1/2}&=&\frac{C(\nu)}{\left(\nu-\frac{1}{2}
\right)}~\frac{H}{2\pi}~\left(-k\eta\right)^{\frac{3}{2}-\nu}, \\
{\cal P}_{\delta\sigma_i}^{~~1/2}&=&\frac{C(\lambda)}
{\left(\nu-\frac{1}{2}\right)}~\frac{H}{2\pi}~
\left(-k\eta\right)^{\frac{3}{2}-\lambda}, 
\end{eqnarray}
where we have used $\eta=-(\nu-1/2)/(aH)$ and we define
\begin{equation}
C(\alpha) \equiv
\frac{2^{\alpha}\Gamma(\alpha)}{2^{\frac{3}{2}}\Gamma(\frac{3}{2})}.
\end{equation}
Both the weighted mean field $\bar\phi$ and the orthonormal fields
$\bar\sigma_i$ are ``light'' fields ($m^2<3H^2/2$) during assisted
inflation (for $\bar{p}>3$) and thus we obtain a spectrum of
fluctuations in all the fields on large scales.
Note that in the de Sitter limit, $\bar{p}\to\infty$ and thus $\nu \to
3/2$ and $\lambda\to3/2$, we have ${\cal P}_{\delta\phi}^{~1/2} \to
H/2\pi$, and ${\cal P}_{\delta\sigma_i}^{~1/2} \to H/2\pi$.

At late times, that is $k\eta\to0$, the perturbations in the weighted
mean field, $\delta\bar\phi$, approach a constant, while the
perturbations in the orthonormal fields, $\delta\bar\sigma_i$, decay
in agreement with our solutions for the homogeneous perturbations
given by Eqs.~(\ref{sigmat}).

Denoting the scale dependence of the perturbation spectra by
$\Delta n_x \equiv d \ln {\cal P}_x /d \ln k$, we obtain
\begin{eqnarray}
\label{nphi}
\Delta n_{\delta\phi}&=&3-2\nu=-\frac{2}{\bar p - 1}, \\
\Delta n_{\delta\bar\sigma_i}&=&3-2\lambda=3 \left( 
1-{\sqrt{(\bar{p}-3)(\bar{p}-1/3)} \over \bar{p}-1} \right). 
\end{eqnarray}


\section{Discussion}
\label{sectdisc}

We have shown that the recently proposed model of assisted inflation,
driven by many scalar fields with steep exponential potentials, can be
better understood by performing a rotation in field space, which allows
us to re-write the potential as a product of a single exponential
potential for a weighted mean field, $\bar\phi$, and a potential $\bar{V}_n$
for the orthogonal degrees of freedom, $\bar\sigma_i$, which has a
global minimum when $\bar\sigma_i=0$. This proves that the scaling
solution found in Ref.~\cite{LMS} is indeed the late-time attractor.

The particular form of the potential which we present for scalar
fields minimally-coupled to the spacetime metric, can also be obtained
via a conformal transformation of a hybrid inflation type inflationary
potential~\cite{hybrid1,hybrid2} with a non-minimally coupled, but
otherwise massless, dilaton field, $\Phi\propto\bar\phi$. Thus we see
that assisted inflation can be understood as a form of power-law
hybrid inflation, where the false-vacuum energy density is diluted by
the evolution of the dilaton field.

We have also been able to give exact solutions for inhomogeneous
linear perturbations about the attractor trajectory in terms of our
rotated fields.
Perturbations in the weighted mean field $\bar\phi$ corresponds to the 
perturbations in the density on the uniform curvature hypersurfaces,
or equivalently, perturbations in the curvature of constant
density hypersurfaces:
\begin{equation}
\zeta = {H \delta\bar\phi \over \dot{\bar\phi}} \, .
\end{equation}
These perturbations are along the attractor trajectory, and hence
describe adiabatic curvature perturbations.
The spectral index of the curvature perturbations on large scales is
thus given from Eq.~(\ref{nphi}) as
\begin{equation}
n_s \equiv 1 + {d \ln {\cal P}_\zeta \over d\ln k} = 1 - {2\over
\bar{p}-1} \,,
\end{equation}
and is always negatively tilted with respect to the
Harrison-Zel'dovich spectrum where $n_s=1$. Note that in the slow-roll
limit ($\bar{p}\to\infty$) we recover the result of Ref.~\cite{LMS}.

First-order perturbations in the fields orthogonal to the weighted
mean field are isocurvature perturbations during inflation. 
Vacuum fluctuations lead to a positively tilted spectrum.
The presence of non-adiabatic perturbations can lead to 
more complicated evolution of the large-scale curvature perturbation than
may be assumed in single-field inflation models~\cite{Salo,SS,twofield}.
However, we have shown that these perturbations decay relative to the
adiabatic perturbations and hence we recover the single field limit at
late times. In particular we find that the curvature perturbation
$\zeta$ becomes constant on super-horizon scales during inflation. Note, 
however, that assisted inflation must be ended by a phase
transition whose properties are not specified in the model. If this 
phase transition is sensitive to the isocurvature (non-adiabatic)
fluctuations orthogonal to the attractor trajectory, then the
curvature perturbation, $\zeta$, during the subsequent radiation
dominated era may not be simply related to the curvature perturbation
during inflation.

\acknowledgments
The authors would like to thank Andrew Liddle and Anupam Mazumdar for
useful discussions.


\end{document}